\documentclass[twoside]{LCWS11}
\usepackage[latin1]{inputenc}
\usepackage[dvips]{graphicx,epsfig,color}
\usepackage{wrapfig,rotating}
\usepackage{amssymb,amsmath,array}
\usepackage{cite}
\newcommand{\geant}         {{\sc Geant4}}
\newcommand{\whizard}      {{\sc Whizard}}
\newcommand{\mokka}      {{\sc Mokka}}
\newcommand{\guinea}      {{\sc Guinea Pig}}
\newcommand{\pandora}      {{\sc Pandora PFA}}

\newcommand{\ra}            {\ensuremath{ \rightarrow     }}

\newcommand{\ep}{e^{+}e^{-}}
\newcommand{\ccg}{\chi\chi\gamma}
\newcommand{\nng}{\nu\nu\gamma}

\newcommand{\lum}{\mathcal{L}}
\newcommand{\Pe}{P_{e^{-}}}
\newcommand{\Pp}{P_{e^{+}}}
\newcommand{\fb}{{\rm fb}^{-1}}

\newcommand{\equal}{{\bf "Equal"}}
\newcommand{\hel}{{\bf "Helicity"}}
\newcommand{\anti}{{\bf "Anti-SM"}}

\definecolor{rred}     {rgb}{1.00, 0.00, 0.00}  
\begin{document}
\title{
Model-independent WIMP Characterisation using ISR} 
\author{Christoph Bartels$^{1}$, Olaf Kittel$^{2}$, 
Ulrich Langenfeld$^{3}$\thanks{U. L. has been supported by funding from the research 
training group GRK 1147 of the Deutsche Forschungsgemeinschaft and 
partially by the Helmholtz alliance 'Physics at the Terascale.}, 
and Jenny List$^{1}$
\thanks{The authors acknowledge the financial support of 
the Deutsche Forschungsgemeinschaft in the DFG project Li~1560/1-1.}
\vspace{.3cm}\\
1- DESY, Notkestrasse 85, D-22607 Hamburg - Germany
\vspace{.1cm}\\
2- Departamento de F\'isica Te\'orica y del Cosmos and CAFPE, \\
Universidad de Granada, E-18071 Granada, Spain
\vspace{.1cm}\\
3- Institut f\"ur Theoretische Physik und Astronomie, \\
Universit\"at W\"urzburg, Am Hubland, D-97074 W\"urzburg, Germany
}

\maketitle

\begin{abstract}
The prospects of measuring the parameters
of WIMP dark matter in a model independent way at the International
Linear Collider are investigated. The signal under study is direct WIMP pair production 
with associated initial state radiation $\ep \ra \ccg$.
The analysis is performed in full simulation of the
ILD detector concept. 
With an integrated luminosity of $\lum = 500\;\fb$ and realistic 
beam polarizations the helicity structure
of the WIMP couplings to electrons can be determined, and
the masses and cross sections can be measured to the percent level.
The systematic uncertainties are dominated by the polarization measurement
and the luminosity spectrum.

\end{abstract}

\section{Radiative WIMP production in $e^+e^-$ collisions}

New Weakly Interacting Massive Particles~(WIMPs) with masses in the order of 
$M_{\chi}\sim 100$~GeV are predicted by several extensions to the SM of particle physics.
With their weak strength interactions, these particles are natural candidates for the
observed abundance of cosmological Dark Matter~(DM).

If these particles were produced at colliders, 
they would leave the detector invisibly without any
further interaction.
Their parameters could be
inferred indirectly by the analysis of cascade decays if other new particles 
exist in the kinematically accessible mass range.
Alternatively the direct pair production of WIMPs with associated initial state radiation
$\ep \ra \chi\chi\gamma$ can be employed to determine the WIMP properties 
from the observed photon spectrum.
It has to be noted, that even if the detection via cascade decays
is possible, the single photon plus missing energy signature
provides an additional measurement of the WIMP candidate.


The rate of radiative WIMP production at an $\ep$ collider can be
estimated model-independent without any assumptions on the dynamics of the interaction 
involved~\cite{Birkedal:2004xn}. With only one new stable particle responsible
for the observed DM content in the universe, the production cross section
for WIMP pairs with associated ISR can be written in the limit of non-relativistic
final state WIMPs as:
\begin{equation}
\frac{d\sigma}{dx\; d\cos{\Theta}}\approx \frac{\alpha\kappa_e\sigma_{\rm an}}{16\pi}
\frac{1+(1-x)^2}{x\sin{\Theta}^2}2^{2J_0}(2S_{\chi}+1)^2 \left(1-\frac{4M_{\chi}^2}{(1-x)s}\right)^{1/2+J_0},
\label{Eqn:WIMP}
\end{equation}
with the candidate mass $M_{\chi}$, the candidate spin $S_{\chi}$ and the center-of-mass energy squared
$s$. The double differential cross section is expressed in the dimensionless variables
$x=\frac{2E_{\gamma}}{\sqrt{s}}$ and $\Theta$ of the emitted photon. 
In Equation~\ref{Eqn:WIMP}, $J_{0}$ is the quantum number of the dominant partial wave
in the production process. In the following, we will refer to the
cases of $J_{0} =0$ (s-wave)  and $J_{0} =1$ (p-wave) production only.
The quantitiy $\kappa_{e}$ is the ``annihilation fraction'' of WIMPs into
electrons.
The parameter $\sigma_{an}$ provides the
overall scale of the production cross section and can be inferred from
observation when the WIMP is identified with the cosmological Dark Matter~\cite{Birkedal:2004xn}.

The annihilation fraction $\kappa_{e}$ implicitely depends on the
helicity of the initial state electrons. For our analysis we investigated
the following three coupling scenarios\cite{DESY-THESIS-2011-034}:
\begin{itemize}
 \item \equal: The WIMP couplings are independent of the
       helicity of the incoming electrons and positrons, 
       i.e.~$\kappa(e^{-}_R,e^{+}_L) = \kappa(e^{-}_R,e^{+}_R) = \kappa(e^{-}_L,e^{+}_L) = \kappa(e^{-}_L,e^{+}_R)$.
 \item \hel: The couplings conserve helicity and parity,\\
       $\kappa(e^{-}_R,e^{+}_L)  = \kappa(e^{-}_L,e^{+}_R)$; $\;\;\;\kappa(e^{-}_R,e^{+}_R) = \kappa(e^{-}_L,e^{+}_L) = 0$.
 \item \anti: This scenario is a "best case" scenario, since
       the WIMPs couple only to right-handed electrons and left-handed positrons:
       $\kappa(e^{-}_R,e^{+}_L)$.
 \end{itemize}

This ``single photon plus missing energy signature'' has 
Standard Model~(SM) processes with large cross sections for background, the dominant one being
radiative neutrino production $\ep \ra\nng$, which
proceeds for high center-of-mass energies primarily via t-channel
$W$ exchange and hence is strongly polarisation dependent. In addition
other SM processes like radiative Bhabha scattering $\ep\ra\ep\gamma$
or multi-photon final states can mimic the WIMP production signature when the
accompanying electrons or photons leave the detector through 
the beam pipe, or are not properly reconstructed. Polarised beams can be used
to significantly reduce these backgrounds and increase the
S/B ratio\cite{MoortgatPick:2005cw},  but the still large abundance of
background events requires a consideration of sytematic uncertainties from the detector measurement and
beam parameters.

\section{Data samples and event selection}

To cover a broad range of parameters in terms of
candidate masses, partial waves and coupling structures, 
only the SM background has been generated and simulated explicitly.
The signal contribution to the data is obtained by reweighting the
irreducible SM $\nng$ background, which is indistinguishable from the signal 
on an event-by-event basis, with the cross section ratio $w_{sig}(E_{\gamma}) =\frac{d\sigma(\ccg)}{dE_{\gamma}}/\frac{d\sigma(\nng)}{dE_{\gamma}}$ 
of radiative WIMP and neutrino pair
production. The weights are evaluated in terms of the MC photon energy.

The measurement of the WIMP parameters requires a precise prediction
of the SM background photon distributions. The expectation is
generated by a parametrization of an independent background subsample,
by succesively correcting the SM prediction
for the detector energy resolution and selection efficiencies.
Remaining differences to the simulated detector output
from the beam energy spectrum and unaccounted detector and
reconstruction effects are parametrized with a higher order polynominal.
From the parametrization of the $\nng$ spectra, the signal prediction
is generated\cite{DESY-THESIS-2011-034}.

\subsection{Simulation and reconstruction}
The background events have been generated using \whizard~\cite{Kilian:2007gr}. The beam
energy spectrum was provided by  \guinea~\cite{GuineaPig} for the nominal RDR baseline
parameter set for a $\sqrt{s} = 500$~GeV machine\cite{Phinney:2007zz}. 
The simulation of the detector response was done for the ILD detector concept~\cite{:2010zzd} with the \geant~\cite{Agostinelli:2002hh} based
simulation software \mokka~\cite{Mokka}. Three statistically independent event samples
of $\sim 50\;\fb$ each have been generated and simulated for the dominant
irreducible $\nng$ background.
The first two samples are used for the background and signal contribution to the data,
while the third sample serves as basis for the spectrum parametrization.
In addition data samples with multi-photon final states $\ep
\ra \gamma\gamma (N) \gamma$ and the Bhabha background
$\ep\ra\ep\gamma$ have been simulated.
After event reconstruction using the \pandora~particle flow algorithm~\cite{Thomson:2009rp}
two corrections to the reconstructed events have been applied. The
reconstruction of high energy photons often results in several
additional detections of lower energy photons, because the
clustering stage of the \pandora~algorithm tends to split 
large energy depositions into several smaller distrinct electromagnetic clusters.
To counter the effect of the
cluster fracturing, photon candidates are merged with a cone based
method.
Second, the photon candidate energies have been recalibrated 
with a calibration function accounting for polar angle dependent
fractional energy losses caused by the segmentation of the ILD calorimeter
system.

\subsection{Event selection}\label{sec:selection}
\begin{figure}[htb]
\setlength{\unitlength}{1.0cm}
  \begin{picture}(14.0, 6.5)
    \put( 0.00, 0.00)  {\epsfig{file=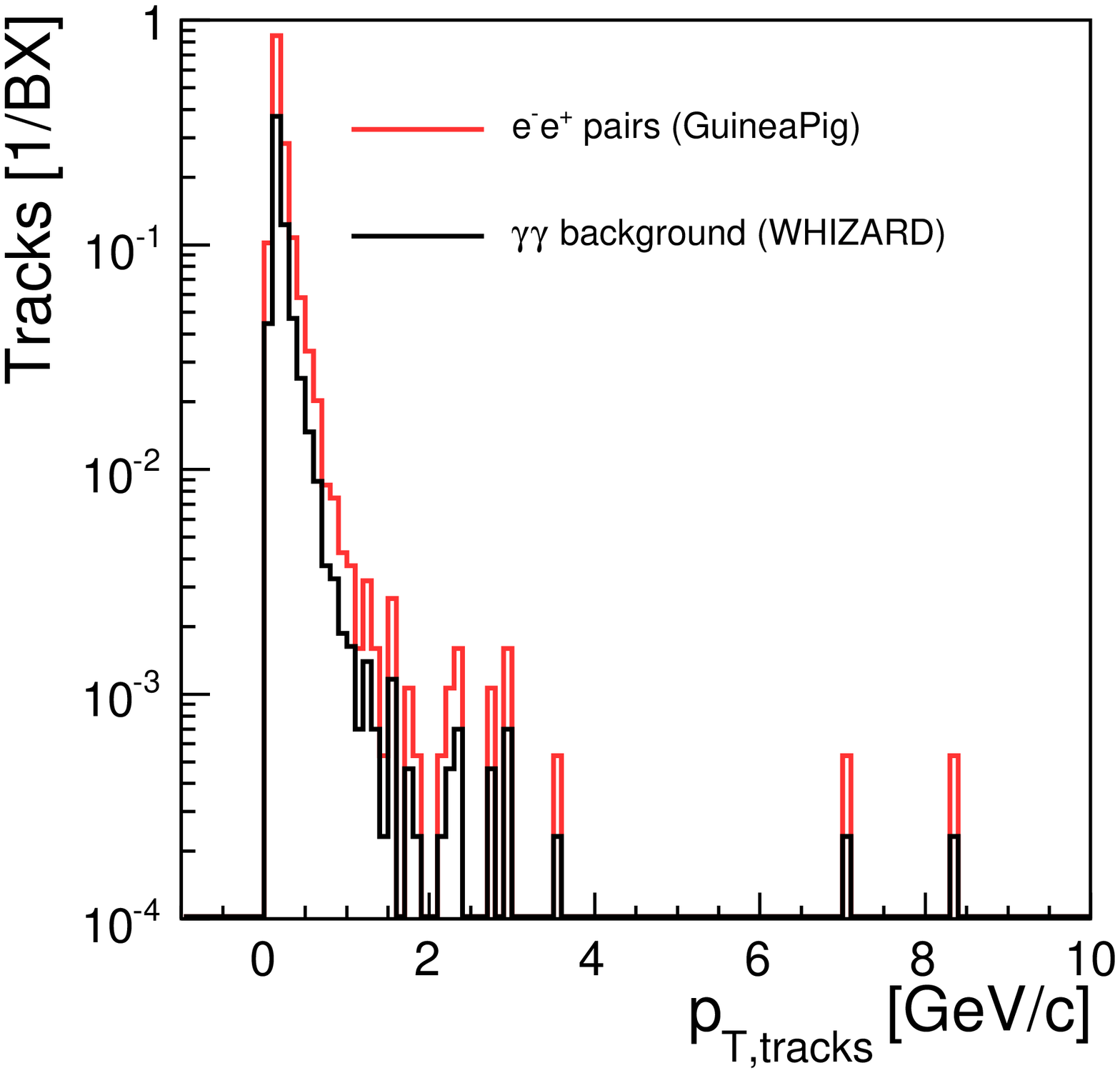, scale=0.35}}
    \put( 7.00, 0.00)  {\epsfig{file=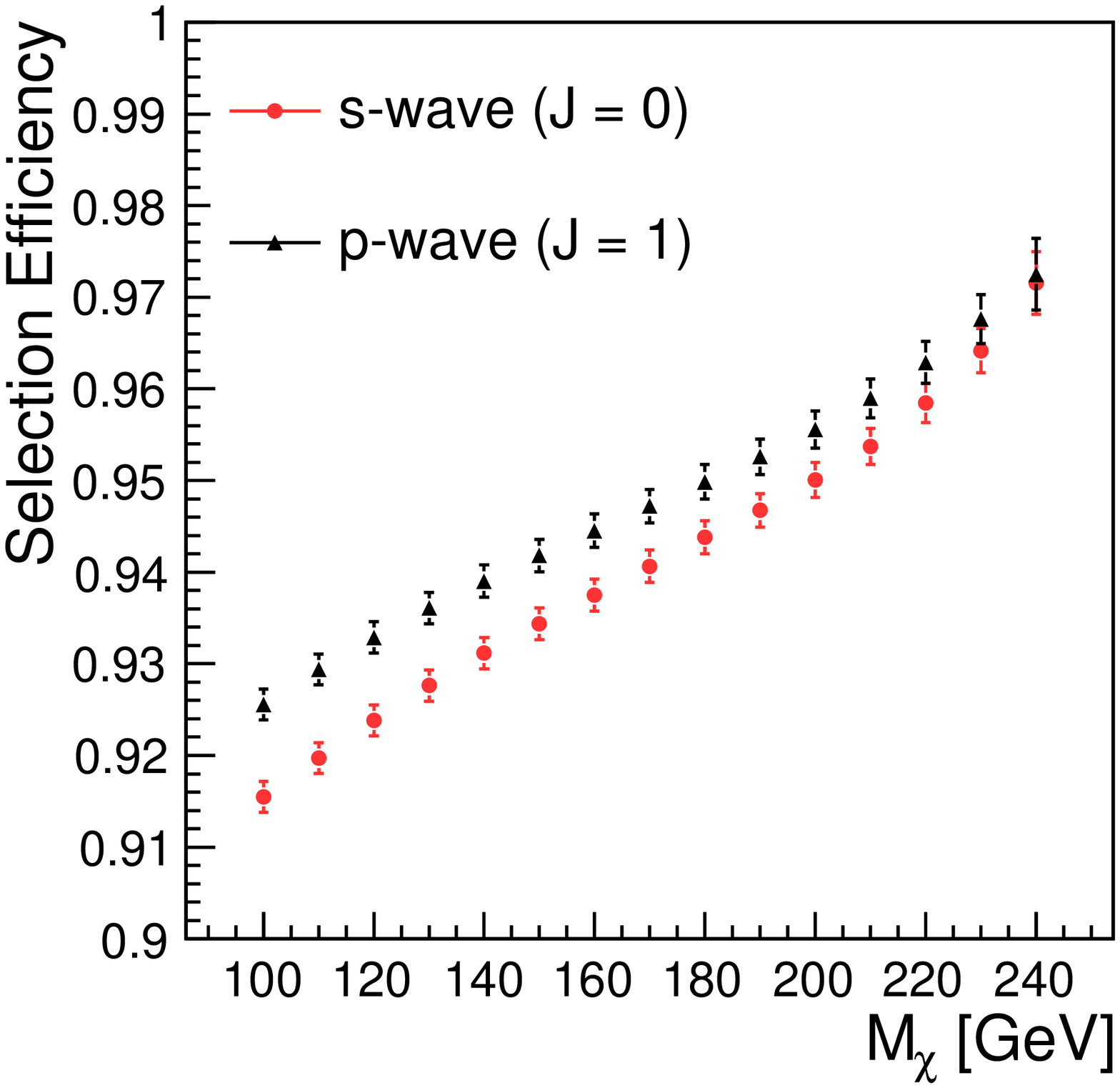, scale=0.35}}
    \put( 0.00, 0.40)  {(a)}
    \put( 8.00, 0.40)  {(b)}
  \end{picture}
  \caption{\label{fig:ptbackground} (a) Transverse momentum distribution of track overlays
from the beamstrahlungs background (red/gray) and from $\gamma\gamma$ processes (black).
(b) Mass dependent signal selection efficiencies for s-wave (red/gray) and p-wave (black)
WIMP production.}
\end{figure}

An event is considered signal-like, if it contains at least one high
$p_{T}$ photon with an energy between 10~GeV $<E_{\gamma}<$ 220~GeV,
and a polar angle constrained to $|\cos{\Theta}| < 0.98$. The condition
on the photon energy reduces the abundant low energy ISR from the SM
background and excludes the massless neutrino final state on the
radiative $Z$ return at photon energies of $E_{\gamma} \approx
241$~GeV. For further event selection, additional constraints are set
to deal with the dominant reducible SM backgrounds, especially radiative Bhabha
scattering and multi-photon final states. To exclude hadronic and
leptonic final states, the maximal exclusive energy
$E_{\gamma}-E_{vis}$, i.e.~the full visible energy excluding the selected photon,
is constrained to 20~GeV.
For further reduction of hadronic final states and Bhabha events, the
maximal allowed transverse track momentum is $p_{T} < 3$~GeV. The
track momenta can not be constrained stronger, as the event selection
has to allow for tracks of $\ep$ pairs from the beamstrahlung
background and for track overlays of multi-peripheral 
$\gamma\gamma\ra$~hadrons events~(collectively called $\gamma\gamma$ processes).
In Figure~\ref{fig:ptbackground}(a) the momentum distribution of tracks
per bunch crossing from the  $\gamma\gamma$ and beamstrahlung background are shown.
On average 0.7 tracks from $\gamma\gamma$ processes and 1.5 tracks
from beamstrahlung are expected in each event. The distributions show a strong
peak at low $p_{T}$ determined from the minimal momentum
for the tracks to reach the tracking region of ILD and fall
of rapidly. Only 0.2\% of tracks have a transverse momentum
above 3~GeV. 
The selection efficieny of the $\nng$ background is on average above 85\%,
with higher efficiencies in the low energy part of the
photon spectrum, where most of the WIMP production signal is located.
This higher efficiency for low energies translates to a signal selection efficiency well above 90\% over the
full mass range, see Figure~\ref{fig:ptbackground}(b). 
\subsection{Systematic uncertainties}

While the WIMP mass is predominantly determined from
the polarisation independent threshold in the photon
energy spectrum, the cross section determination
requires a precise knowledge of the
normalisation of the polarisation dependent
$\nng$ background. The polarisation measurement
precision is assumed to $\delta P/P = 0.25\%$\cite{bib:EP-paper}. 
The precision on the luminosity is given
by the RDR~\cite{Phinney:2007zz} with $10^{-4}$.
A further source  of systematic error is 
given by the knowledge of the beam energy 
spectrum. 
Here, its influence is conservately estimated
from the impact on the signal spectrum for 
two different parameter sets, namely the nominal RDR and SB-2009~\cite{Berggren:2010wy} sets. 
For the cross section measurement which is not sensitive
to the partial wave quantum number, the selection
efficiency obtains an additional contribution
of uncertainty from the difference between the selection
efficiencies of s- and p-wave production. 

\section{Results}

For the analysis a typical running scenario of the ILC is assumed,
where an integrated luminosity of $\lum = 500\;\fb$ is distributed
to four polarisation states with $(+|\Pe|;-|\Pp|)$, $(-|\Pe|;+|\Pp|)$, $(+|\Pe|;+|\Pp|)$
and $(-|\Pe|;-|\Pp|)$. The odd (equal) sign configurations 
obtain $40\%$~($10\%$) of the delivered luminosity each.
The absolute polarization values are assumed to be $|\Pe|=0.8$ and $|\Pp| =0.3$
or $|\Pp| =0.6$, respectively.
The total unpolarised signal cross section in the signal region is
set to $\sigma_{0} = 100$~fb throughout. 
For the determination of the cross section and coupling structure
the candidate mass is fixed to 150~GeV.

\subsection{Helicity structure and cross section}

With four different longitudinal polarization configurations, the fully polarized cross sections 
$\sigma_{\lbrace L,R \rbrace}$, and hence the helicity structure 
of the WIMP couplings to the beam electrons, can be  determined from
the cross section deconstruction~\cite{MoortgatPick:2005cw}
\begin{eqnarray}
\sigma( P_{e^-}, P_{e^+}) & = &
\frac{1}{4}\Big[ (1+P_{e^-})(1+P_{e^+}) \sigma_{RR} 
                  + (1-P_{e^-})(1-P_{e^+}) \sigma_{LL}\nonumber\\
        && + \,\,\, (1+P_{e^-})(1-P_{e^+}) \sigma_{RL} 
                  + (1-P_{e^-})(1+P_{e^+}) \sigma_{LR} \Big].
\label{Eqn:Deconstruct1}
\end{eqnarray}

\begin{table}[!h]
  \centering
  \renewcommand{\arraystretch}{1.08}
  \begin{tabular*}{\textwidth}{l@{\extracolsep{\fill}} p{2mm} rr}  
    \hline\hline 
    \multicolumn{4}{c}{\quad } \\[-4.9mm]
    \quad                    && $(|\Pe|;|\Pp|) = (0.8;0.3)$ & $(|\Pe|;|\Pp|) = (0.8;0.6)$ \\[0.5pt]
    \hline\hline 
    \multicolumn{4}{c}{\quad} \\[-2mm]
    \multicolumn{4}{l}{\quad{\bf\hel}\quad scenario } \\ \hline
    \multicolumn{4}{c}{\quad} \\[-4.9mm]
    $\sigma_{RL}/\sigma_{0}$ &&    $1.99 \pm 0.24$ \quad $(0.16)$  &   $ 1.99 \pm 0.10$ \quad $(0.08)$ \\
    $\sigma_{RR}/\sigma_{0}$ &&    $0.00 \pm 0.33$ \quad $(0.21)$  &   $ 0.00 \pm 0.23$ \quad $(0.14)$ \\
    $\sigma_{LL}/\sigma_{0}$ &&    $0.00 \pm 0.37$ \quad $(0.29)$  &   $ 0.00 \pm 0.23$ \quad $(0.15)$ \\
    $\sigma_{LR}/\sigma_{0}$ &&    $1.95 \pm 0.38$ \quad $(0.25)$  &   $ 1.95 \pm 0.29$ \quad $(0.16)$ \\
    \hline 
  \end{tabular*}
  \caption[Measurement results of fully polarized cross sections] {
    Fully polarized cross sections $\sigma_{\lbrace R,L\rbrace}$ measured within 
    the {\bf\hel} WIMP scenario and for two different absolute polarizations of electrons and positrons. 
    The quoted uncertainties are the squared sum of statistical errors and systematic uncertainties, 
    with the bracketed values corresponding to an increased precision 
    on the polarization measurement of $\delta P/P = 0.1\%$.
  }
  \label{Table:XSecPolMeas1}
\end{table}
\begin{figure}[htb]
\setlength{\unitlength}{1.0cm}
  \begin{picture}(14.0,6.5)
  \put(0.0,0.0){\epsfig{file=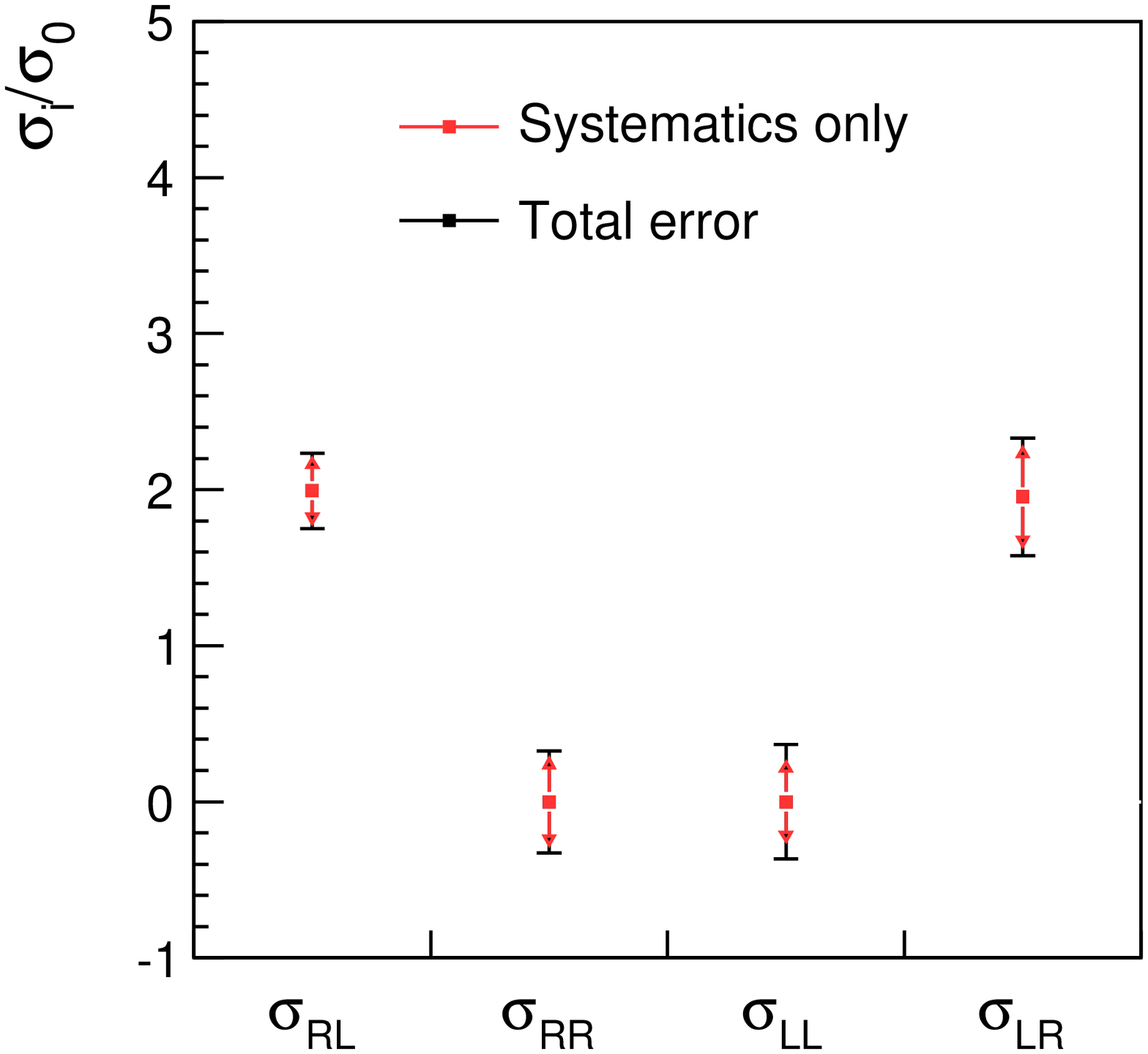,scale=.35}} 
  \put(7.0,0.0){\epsfig{file=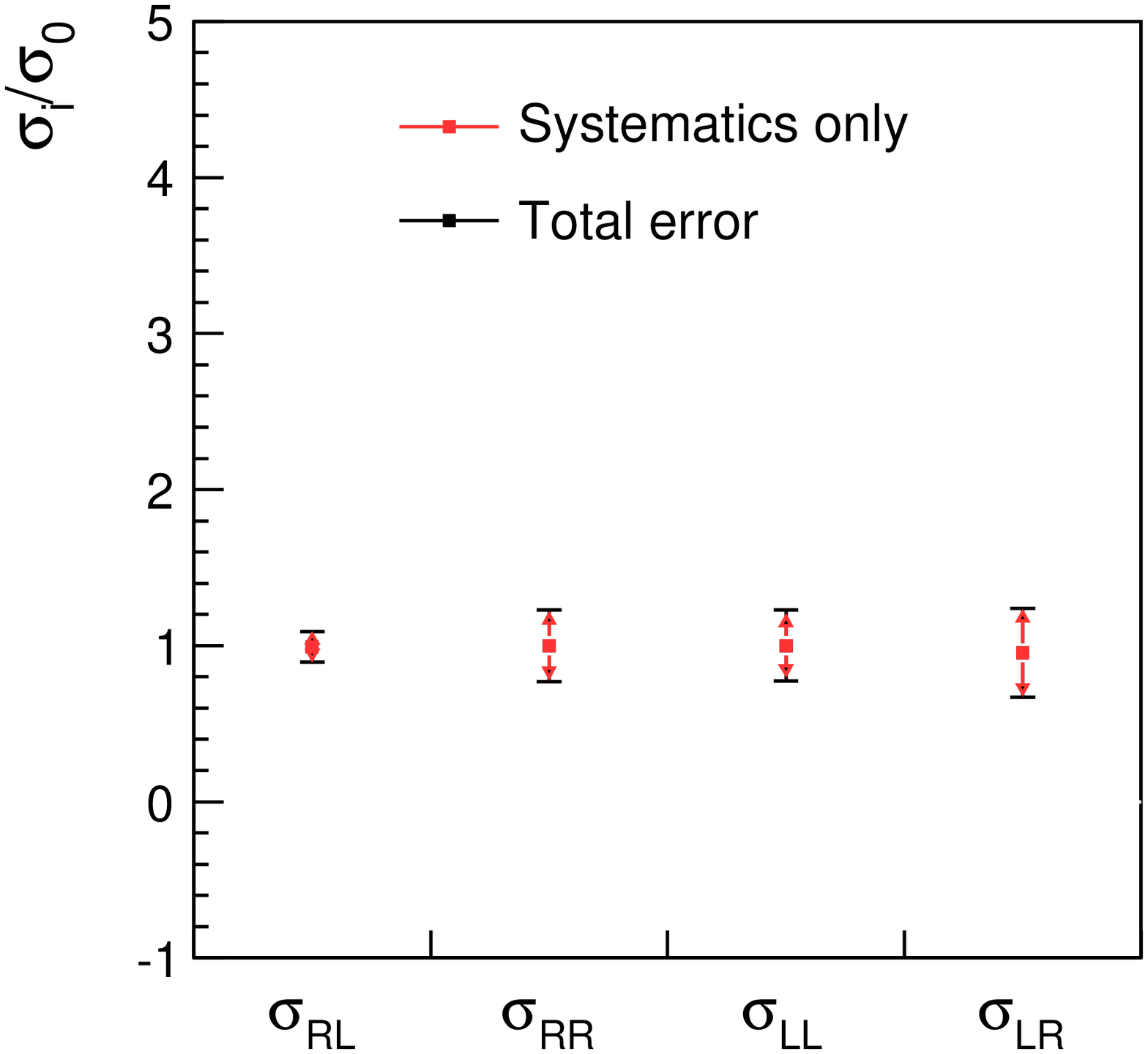,scale=.35}}
  \put(0.0,0.0){(a)}   
  \put(7.0,0.0){(b)} 

  \end{picture}
  \caption{\label{fig:polxsect} Helicity structure of WIMP couplings in terms of the fully
polarised cross sections  $\sigma_{\lbrace R,L\rbrace}$ in (a) the \hel{} scenario and (b) the \equal{} scenario.
The values are normalised to the unpolarised cross section $\sigma_{0}$.}
\end{figure}
The results for the \hel{} coupling scenario are listed in terms of the fully polarized
cross sections $\sigma_{\lbrace R,L\rbrace}$ in Table~\ref{Table:XSecPolMeas1} and shown
in Figure~\ref{fig:polxsect}(a). Figure~\ref{fig:polxsect}(b) depicts 
the mesurement in the \equal{} scenario. 
Given the input cross section of $100$~fb, the fully
polarized cross sections can be determined to
20~fb to 40~fb with a positron polarization of 30\%.
The uncertainties are reduced to 10 to 30~fb with an 
increased positron polarization of 60\%.
This is in particular the case for $\sigma_{\lbrace L,R \rbrace}$,
wich is primarily determined by the measurement
$\sigma_{+-}$ with positive electron and negative positron
polarization. In that case the SM background is maximally reduced.
The dominant source of systematic uncertaity
stems from the polarization measurement.

Assuming uncorrelated polarisation errors, a combination of the four individual measurements provides
a determination of the unpolarised cross section $\sigma_{0}$
to a precision of 3 to 5\% for an positron polarisation 
of $P = 30\%$, depending on the coupling scenario, see Table~\ref{Table:XSecUnPolMeas}. 
With an increased positron polarisation
of $P = 60\%$, the relative precision is increased to 2.5\% . 
Again the dominant source of systematic uncertainty
comes from the polarization measurement.
\begin{table}[!h]
  \centering
  \renewcommand{\arraystretch}{1.10}
  \begin{tabular*}{\textwidth}{l@{\extracolsep{\fill}} p{6mm} rl p{4mm} rl}  
    \hline\hline 
    \multicolumn{7}{c}{\quad } \\[-4.8mm]
    Data scenario && \multicolumn{5}{c}{Unpolarized cross section: $\;\sigma_{0} \;\pm\; {\rm stat} \;\pm\,{\rm sys}$\quad $(\pm\,{\rm total})$ [fb]} \\
    (simulated)   && \multicolumn{2}{c}{$(|\Pe|;\,|\Pp|)\;=\;(0.8;\,0.3)$} &&\multicolumn{2}{c}{$(|\Pe|;\,|\Pp|)\;=\;(0.8;\,0.6)$} \\[1pt]
    \hline\hline 
    \multicolumn{7}{c}{\quad } \\[-2mm]
    \multicolumn{7}{l}{ Assumed polarization uncertainty\quad $\delta P/P = 0.25\%$ } \\ \hline
    \multicolumn{7}{c}{\quad } \\[-4.8mm]
    \equal &&    $99.0 \;\pm\; 2.8 \;\pm\; 4.3$  &  $(\pm\; 5.1)$   &&    $99.2 \;\pm\; 2.7 \;\pm\; 3.5$  &  $(\pm\; 4.4)$ \\
    \hel   &&    $99.1 \;\pm\; 2.3 \;\pm\; 4.0$  &  $(\pm\; 4.6)$   &&    $99.4 \;\pm\; 2.0 \;\pm\; 2.8$  &  $(\pm\; 3.4)$ \\
    \anti  &&    $99.8 \;\pm\; 1.4 \;\pm\; 2.8$  &  $(\pm\; 3.2)$   &&    $99.7 \;\pm\; 1.1 \;\pm\; 2.1$  &  $(\pm\; 2.4)$ \\
    \hline 
  \end{tabular*}
  \caption[Measurement results of unpolarized cross section in three studied scenarios]{
    Measured unpolarized cross section $\sigma_{0}$ by a combination of cross section measurements 
    with polarized beams for an integrated luminosity of $\lum =500\;\fb$.}
  \label{Table:XSecUnPolMeas}
\end{table}

\subsection{Mass measurement and partial wave}
The candidate mass is measured by fitting template spectra for s+b
to the measured data spectrum. The mass is determined
by the $\chi^{2}$ of the measurement. Depending on the 
coupling scenario, polarization configuration, the candidate mass
can be determined to a level of $< 2\%$, see Table~\ref{Table:MassMeasuremnt}.

\begin{table}[!h]
  \centering
  \renewcommand{\arraystretch}{1.10}
  \begin{tabular*}{\textwidth}{l@{\extracolsep{\fill}} rrr}  
    \hline\hline 
    \multicolumn{4}{c}{\quad } \\[-4.8mm]
    Mass   & \multicolumn{3}{c}{WIMP mass: $\;\pm\,$stat.  $\pm\,\delta E\,$(sys.)  $\pm\,\delta\lum\,$(sys.)$\;$ (total) [GeV]} \\
    \protect[GeV\protect]\hspace*{5mm}
    $\,$   &     $(\Pe;\,\Pp)\,=\,(0.8;\, 0.0)$  &     $(\Pe;\,\Pp)\,=\,(0.8;\,-0.3)$  & $(\Pe;\,\Pp)\,=\,(0.8;\,-0.6)$ \\[1pt]
    \hline\hline 
    \multicolumn{4}{c}{\quad} \\[-2.5mm]
    \multicolumn{4}{l}{\quad\hel\quad scenario } \\ \hline
    \multicolumn{4}{c}{\quad} \\[-4.8mm]
    $120$  &  $2.67 \pm0.07 \pm1.91\;\; (3.29)$  &  $1.92 \pm0.07 \pm1.89\;\; (2.70)$  &  $1.53 \pm0.07 \pm1.89\;\; (2.43)$ \\
    $150$  &  $2.11 \pm0.05 \pm1.47\;\; (2.57)$  &  $1.62 \pm0.05 \pm1.46\;\; (2.18)$  &  $1.23 \pm0.05 \pm1.45\;\; (1.90)$ \\
    $180$  &  $1.78 \pm0.03 \pm1.00\;\; (2.04)$  &  $1.36 \pm0.03 \pm1.00\;\; (1.69)$  &  $0.94 \pm0.03 \pm1.00\;\; (1.37)$ \\
    $210$  &  $0.78 \pm0.02 \pm0.54\;\; (0.95)$  &  $0.67 \pm0.02 \pm0.54\;\; (0.87)$  &  $0.59 \pm0.02 \pm0.54\;\; (0.80)$ \\
    \hline 
  \end{tabular*}
  \caption[Errors on measured masses for all three coupling scenarios and three polarization configurations.]{
    Statistical and systematic uncertainties on the measured WIMP masses 
    for an integrated luminosity of $\lum = 500\;\fb$ 
    in the \hel{} coupling scenarios for three different polarization configurations.}
  \label{Table:MassMeasuremnt}
\end{table}

The dominant source of systematic uncertainty come from the
calibration of the overall energy scale ($\delta E$) and the
measurement of the beam energy spectrum ($\delta \lum$).

An indication of the dominant partial wave in the production process
is obtained from the $\chi^{2}$ value of fitting s- and p-wave
template spectra against the data spectrum.
In all studied scenarios the
fit converges better for the correct partial wave assumption.
However, with unpolarised beams, the distinction is not very
clear over the full mass range. Utilising the possibility
of polarised positrons allows to clearly seperate the s- and p-wave production.

\begin{footnotesize}


\end{footnotesize}

\clearpage


\end{document}